\documentclass[9pt,letterpaper,twocolumn,pra,showpacs]{revtex4-1}
\usepackage[utf8]{inputenc}
\usepackage{amsmath}
\usepackage{mathptmx}
\usepackage{graphicx}

\begin{document}
\title{Modulational Instability of Nonlinear Polarization Mode Coupling in Microresonators}

\author{T. Hansson}
\email{Corresponding author: tobias.hansson@unibs.it}
\affiliation{Dipartimento di Ingegneria dell'Informazione, Universit\`a di Brescia, via Branze 38, 25123 Brescia, Italy}

\author{M. Bernard}
\affiliation{Dipartimento di Ingegneria dell'Informazione, Universit\`a di Brescia, via Branze 38, 25123 Brescia, Italy}

\author{S. Wabnitz}
\affiliation{Dipartimento di Ingegneria dell'Informazione, Universit\`a di Brescia, via Branze 38, 25123 Brescia, Italy}
\affiliation{CNR–INO, via Branze 38, 25123 Brescia, Italy}

\begin{abstract}
We investigate frequency comb generation in the presence of polarization effects induced by nonlinear mode coupling in microresonator devices. A set of coupled temporal Lugiato-Lefever equations are derived to model the propagation dynamics, and an in-depth study is made of the modulational instability of their multistable homogeneous steady-state solutions. It is shown that new kinds of instabilities can occur for co-propagating fields that interact through nonlinear cross-phase modulation. These instabilities display properties that differ from their scalar counterpart, and are shown to result in the generation of new types of incoherently coupled frequency comb states.
\end{abstract}

\maketitle

\section{Introduction}

Optical frequency combs are currently attracting significant research interest due to their many application for areas such as spectroscopy, metrology and quantum optics \cite{Kippenberg_2011,Pasquazi_2017}. Especially interesting is the experimental approach based on continuous wave (CW) driven microresonators, that can produce wide bandwidth combs with high repetition rates in a compact device by the resonant enhancement of nonlinear interactions that occur in a small modal volume.

The mechanism for the generation of frequency combs in Kerr microresonators is fundamentally dependent on the instability of the CW intracavity field to periodic modulations. In particular, for resonators that are driven at a single frequency, it is the vacuum noise seeded modulational instability (MI) that is responsible for the initial generation of sidebands that grow around the pump wavelength \cite{Matsko_2005,Hansson_2013}. As the sidebands grow, they can in turn interact with the pump and each other through cascaded four-wave mixing (FWM) processes to generate new sidebands that bring about the formation of a wide multi-line frequency comb. As a consequence, the type of comb states that can form inside the resonator is highly influenced by the spectral properties of the MI gain, as well as by the dispersion properties of the material. In dissipative cavities the MI exhibits a power threshold because of the need for the growth rate of the instability to exceed the roundtrip loss. For power levels that are close to this threshold, it is generally possible to find stationary and mode-locked comb states such as periodic (Turing) patterns or cavity solitons \cite{Coillet_2013}, that are dissipative extensions of cnoidal waves \cite{Qi_2017}, and which remain in a stable equilibrium due to saturation of the instability growth rate through the continuous redistribution of energy to distant sidebands.

To gain insight into these processes, it is important to develop an understanding of the comb formation dynamics. For the case when the field propagates in a single mode family, the comb dynamics can, in the mean-field approximation, be modelled by the scalar Lugiato-Lefever equation (LLE) \cite{LL_1987,Coen_2013}. The LLE has homogeneous CW solutions that have been shown to be unstable to MI with a growth rate that is dependent on a single real and positive eigenvalue. However, in a more general case when, e.g., the nonlinear phase shift is large \cite{Hansson_2015} or the polarization degree of freedom is taken into account, one finds that other types of instabilities also appear. Since these instabilities are associated with different normal modes they can lead to the formation of entirely new types of comb states with respect to the scalar MI. In fact, previous studies of polarization mode coupling in birefringent optical fibers have shown that nonlinear cross-phase modulation (XPM) allows MI to occur in the normal dispersion regime, with the instability being either parallel (scalar MI) \cite{Agrawal_1987} or orthogonal (polarization MI) \cite{Wabnitz_1988} to the pump field polarization. Moreover, it has been found that MI can also develop with a single frequency shifted sideband in each polarization direction \cite{Rothenberg_1990}. It should be expected that similar instabilities will occur also in polarization mode coupled microresonators, although with properties that are modified by the presence of the cavity boundary conditions.

The various interactions of modes with different order or polarization are of primary importance in microresonator devices. Two resonances, that in the absence of mode coupling would be degenerate in frequency, may shift their respective resonances to avoid a mode crossing \cite{Liu_2014}. This can lead to large modifications of the dispersion around the crossing point that may, e.g., result in a mode family having nominally normal dispersion becoming locally anomalous. Such avoided mode crossings have been identified as the principal mechanism for initiating comb generation in the normal dispersion regime \cite{Savchenkov_2012}, and are also associated with the formation of soliton crystals \cite{Cole_2017}. The nonlinear coupling of resonator modes has been studied theoretically as early as 1994 in the context of fiber ring cavities \cite{Haelterman_1994}, and has predicted the possibility of polarization MI and symmetry breaking. Recent work on microresonators has also shown the possibility of generating trains of polarization locked vector solitons structures \cite{Menyuk_2016,Menyuk_2017}, as well as coexisting multistable soliton states (i.e. super cavity solitons \cite{Hansson_2015}) having different peak power and width for the same pump parameters \cite{Averlant_2017}. Moreover, experiments have demonstrated the induction of a secondary frequency comb in an orthogonal polarization through XPM \cite{Bao_2017}. Compound soliton states that are bound due to birefringence have also recently been observed in fiber ring cavities \cite{Wang_2017}.

Modulational instability in the presence of polarization effects was analyzed in Ref.~\cite{Haelterman_1994} for the case of an isotropic fiber ring cavity. The cavity was assumed to be excited by a linearly polarized pump field, and a set of mean-field equations with symmetric coefficients was derived for the two counter-rotating circularly polarized modes of the fiber. An MI analysis was performed, and conditions were derived for which the MI could occur with perturbations growing either parallel or orthogonal to the steady-state field. In the present article we are considering a related, yet clearly distinct, physical situation, corresponding to a microresonator device with two orthogonally polarized TE/TM modes that can feature different dispersion characteristics. We consider an arbitrary CW pump field excitation of the two modes, and analyze the MI of the steady-state solutions under the more general conditions that are valid for microresonator frequency comb generation. We begin in Section II, by deriving a coupled pair of mean-field equations for modelling the evolution of the two polarization components. The properties of the homogeneous steady-state solutions of these equations are then analyzed in Section III, where we also discuss conditions for their symmetry breaking. In Section IV, we perform an in-depth MI analysis of the steady-state solutions, and consider the explicit solution of the characteristic eigenvalue equation for two complementary limiting cases. We derive simplified phase-matching conditions that can be used to predict the occurrence of MI, and give examples for which different types of instabilities are observed. In Section V, we present numerical simulation results of frequency comb states that are generated for the different MIs predicted by our analysis. We conclude the paper in Section VI and discuss the possibility of using the polarization degree of freedom for generating frequency combs in the normal dispersion regime.

\section{Temporal mean-field model}

We consider nonlinear coupling of two orthogonally polarized mode families in a microresonator. In particular, we consider TE/TM modes that preserve their state of polarization in the absence of linear coupling through scattering interactions. Differently from Ref.~\cite{Menyuk_2016}, we make use of a two timescale approach and derive the evolution equations starting from an Ikeda map \cite{Haelterman_1994}.

Inside the resonator waveguide, the slowly varying envelope of the electric field with angular carrier frequency $\omega_0$ is taken to have two linearly polarized vector components $A_j = A_j(z,\tau)$ along the principal axes $j = x,y$, respectively. Assuming the material of the waveguide to be an isotropic Kerr medium, these satisfy the following set of dissipative single pass evolution equations (see, e.g., \cite{Agrawal}) for a given roundtrip $m$

\begin{align}
  & \frac{\partial A^m_x}{\partial z} + \beta_{1,x}\frac{\partial A^m_x}{\partial\tau} + i\frac{\beta_{2,x}}{2}\frac{\partial^2 A^m_x}{\partial\tau^2} + \frac{\alpha_{i,x}}{2}A^m_x = \nonumber\\
  & i\gamma\left(|A^m_x|^2+ \frac{2}{3}|A^m_y|^2\right)A^m_x + i\frac{\gamma}{3}(A^m_x)^*(A^m_y)^2e^{-i2\Delta\beta z},\\
  & \frac{\partial A^m_y}{\partial z} + \beta_{1,y}\frac{\partial A^m_y}{\partial\tau} + i\frac{\beta_{2,y}}{2}\frac{\partial^2 A^m_y}{\partial\tau^2} + \frac{\alpha_{i,y}}{2}A^m_y = \nonumber\\
  & i\gamma\left(|A^m_y|^2+ \frac{2}{3}|A^m_x|^2\right)A^m_y + i\frac{\gamma}{3}(A^m_y)^*(A^m_x)^2e^{i2\Delta\beta z}.
\end{align}
Here $z$ is the longitudinal coordinate and $\tau$ is (fast) time. The dispersion properties are derived from the propagation constant $\beta_j(\omega)$ that has expansion coefficients defined by $\beta_{k,j} = d^k\beta_j/d\omega^k|_{\omega_0}$. The amount of linear birefringence is in particular determined by $\Delta\beta = \beta_{0,x}-\beta_{0,y}$, while $\beta_{1,j}$ and $\beta_{2,j}$ are the inverse group-velocity and group-velocity dispersion coefficients, respectively. The nonlinear coefficient $\gamma = k_0n_2/A_{\textrm{eff}}$ gives the strength of the Kerr nonlinearity, under the assumption of equal mode areas, and the damping coefficients $\alpha_{i,j}$ denote the total linear loss per unit length of the waveguide.

In addition, the fields satisfy cavity boundary conditions that are given by
\begin{equation}
  A^{m+1}_j(z = 0,\tau) = \sqrt{\theta_j}E_j^{in} + \sqrt{1-\theta_j}e^{-i\delta_j}A^m_j(z=L,\tau).
\end{equation}
These close the system by coupling fields between successive roundtrips and model the injection of the external CW pump field with vector components $E_j^{in}$. The amount of power coupled into the resonator is dependent both on the separate coupling coefficients $\theta_j$ and the detunings $\delta_j$ of the pump modes with respect to the pump laser frequency.

In the following, we assume that the fields do not change significantly over the length $L$ of a single roundtrip and perform a mean-field averaging of the map by integrating the field evolution along the resonator circumference. After applying the boundary condition, with the detuning and coupling coefficients taken to be quantities of the first order \cite{Haelterman_1992}, we obtain two coupled mean-field equations that generalize the LLE, viz.
\begin{align}
  & \frac{\partial E_x}{\partial t} = \left[-(\alpha_x+i\delta_x) +\Delta\beta'_1 \frac{\partial}{\partial\tau} - i\frac{\beta'_{2,x}}{2}\frac{\partial^2}{\partial\tau^2}\right]E_x\nonumber\\
  & +i\left(|E_x|^2+ \sigma|E_y|^2\right)E_x + S_x, \label{eq:S1}\\
  & \frac{\partial E_y}{\partial t} = \left[- (\alpha_y+i\delta_y) - \Delta\beta'_1 \frac{\partial}{\partial\tau} - i\frac{\beta'_{2,y}}{2}\frac{\partial^2}{\partial\tau^2} \right]E_y\nonumber\\
  & +i\left(|E_y|^2+ \sigma|E_x|^2\right)E_y + S_y, \label{eq:S2}
\end{align}
where the cross-coupling parameter $\sigma = 2/3$ for linearly polarized light, and we have normalized the fields and introduced a slow-time variable $t$ such that $E_j(t = m,\tau)=A^m_j(z=0,\tau)/\sqrt{\gamma L}$. In addition, we have made a transformation to a reference frame that moves with the mean group-velocity and defined $\beta'_j = \beta_j L$ with the group-velocity mismatch (GVM) $\Delta\beta'_1 = (\beta'_{1,x}-\beta'_{1,y})/2$, the round-trip loss $\alpha_j = (\alpha_{i,j}L+\theta_j)/2$, and the normalized pump field components $S_j = \sqrt{\gamma L\theta_j}E_j^{in}$. We have also neglected coherent FWM terms on account that they generally average out to zero due to the difference in effective refractive index between the two modes, whose respective resonances we assume are associated with different mode numbers. The equations are as a consequence incoherently coupled and the modes are unable to exchange power.

The two mean-field equations above constitute the basic model that will be analyzed in the following. To the first order of approximation, they include the effects of group-velocity mismatch, group-velocity dispersion, self-phase modulation (SPM) and cross-phase modulation as well as damping, driving and detuning of the pump laser frequency.

\section{Homogeneous steady-state solutions}

An initially empty cavity that is being driven by a pump laser will, in the absence of instabilities, converge to a homogeneous steady-state solution that satisfy the following coupled system of algebraic equations
\begin{align}
  & P_x = \left[\left(\delta_x-I_x-\sigma I_y\right)^2+\alpha_x^2\right]I_x, \label{eq:HSx}\\
  & P_y = \left[\left(\delta_y-I_y-\sigma I_x\right)^2+\alpha_y^2\right]I_y, \label{eq:HSy}
\end{align}
where we have defined $P_j = |S_j|^2$ and $I_j = |E_j|^2$. Considering the polarization components independently, Eqs.~(\ref{eq:HSx}-\ref{eq:HSy}) have the standard form of bistable resonances with nonlinear phase shifts depending on both SPM and XPM.

\begin{figure}[htb]
\centerline{\includegraphics[width=1.1\linewidth]{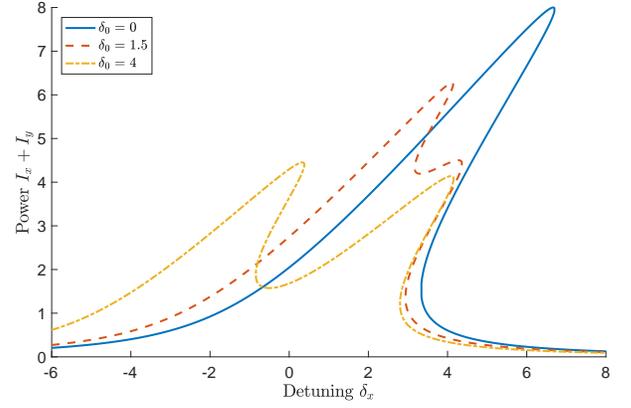}}
\caption{Homogeneous steady-state solutions showing Kerr tilts for $P_x = P_y = 4$, $\sigma = 2/3$ and $\alpha_x = \alpha_y = 1$ with $\delta_0 = 0$ (blue solid), $\delta_0 = 1.5$ (red dashed) and $\delta_0 = 4$ (orange dash-dotted). The intermediate case of $\delta_0 = 1.5$ is seen to display tristability. The positive slope branches on the right hand side of each tilted resonance are unstable to CW perturbations.}
\label{fig:res}
\end{figure}

To better understand the steady-state behaviour, we consider the total field intensity as a function of detuning. Assuming moderate pump power levels and a fixed difference in detuning between the two modes, i.e. $\delta_y = \delta_x+\delta_0$, we find that the total power allows either a single or split resonance shape. In particular, for the degenerate case we have as $\delta_0 \to 0$ a single Kerr tilted resonance with bistable behaviour, or in the opposite limit, two isolated split resonances, see Fig.~\ref{fig:res}. In an intermediate regime, we also find the possibility of split resonances with a tristable behaviour. Such tristable states have been shown to allow for the generation of a multistable pair of cavity solitons with different amplitudes and widths \cite{Averlant_2017}. As usual, the bistability requires a minimum detuning that can be derived from the condition that $dI_x/dP_x = 0$ has real roots for the intracavity power, and that is given by $\delta_x \geq \sqrt{3}\alpha_x + \sigma I_y$ and analogous for the y-component. The power in each component is bound by $0 \leq I_j \leq P_j/\alpha_j^2$, and the components attain their individual maxima at the peak of their respective resonances where the detuning compensates for the nonlinear phase shifts.

The coupled steady-state Eqs.~(\ref{eq:HSx}-\ref{eq:HSy}) may admit up to nine different solutions for a given set of parameters. Among these are symmetry breaking solutions \cite{Haelterman_1994}. Assuming equal detuning, loss and pump power the system reduces to two symmetrical equations for $I_{x,y}$. The difference between these satisfy the consistency requirement
\begin{equation}
  (I_x - I_y)\left[(I_x+I_y)^2 - I_xI_y(1-\sigma)^2 - 2\delta_x(I_x+I_y) + (\alpha^2+\delta_x^2)\right] = 0,
\end{equation}
where the first factor represents the symmetric solution. An asymmetric, symmetry breaking, pair of solutions can also be found by setting the second factor to zero. The bifurcation points for these solutions are given by $I_x = I_y = \left(2\delta_x \pm \sqrt{(1-\sigma)^2\delta_x^2-(4-(1-\sigma)^2)\alpha^2}\right)/(4-(1-\sigma)^2)$, and a condition for their existence is that the detuning must be sufficiently large, $\delta_x \geq \alpha\sqrt{4/(1-\sigma)^2-1} \approx 6\alpha$. However, with $\sigma = 2/3$ these solutions originate on the unstable branch and are generally hard excitations that are unreachable by adiabatic changes to pump power and detuning \cite{Matsko_2012}.

\section{Modulation instability analysis}

To investigate the MI we use an ansatz for the perturbation of the form $E_i = E_{0,i} + u_i$, with $u_i$ complex, and linearize Eqs.~(\ref{eq:S1}-\ref{eq:S2}) around the steady-state solution. To simplify the calculations we make the assumption that each mode has a reference phase such that the CW solutions are purely real. The linearized equation for the Fourier transform can then be written as $d\tilde{\mathbf{u}}/dt = A \tilde{\mathbf{u}}$, with $\tilde{\mathbf{u}} = [\tilde{u}_x,\tilde{u}_x^*,\tilde{u}_y,\tilde{u}_y^*]^T$, and with the coefficient matrix given by
\begin{equation} \scriptsize
  A = \left[\begin{array}{cccc}
      iq_x-\alpha_x-i\Delta\beta'_1\omega & iI_x & i\sigma\sqrt{I_xI_y} & i\sigma\sqrt{I_xI_y} \\
      -iI_x & -iq_x-\alpha_x-i\Delta\beta'_1\omega & -i\sigma\sqrt{I_xI_y} & -i\sigma\sqrt{I_xI_y} \\
      i\sigma\sqrt{I_xI_y} & i\sigma\sqrt{I_xI_y} & iq_y-\alpha_y+i\Delta\beta'_1\omega & iI_y \\
      -i\sigma\sqrt{I_xI_y} & -i\sigma\sqrt{I_xI_y} & -iI_y & -iq_y-\alpha_y+i\Delta\beta'_1\omega
      \end{array}\right],
\end{equation}
where $\omega$ is the modulation frequency, and we have defined the two phase-matching conditions
\begin{equation}
  q_x = 2I_x + \sigma I_y - \delta_x + \frac{\beta'_{2,x}}{2}\omega^2, \qquad q_y = 2I_y + \sigma I_x - \delta_y + \frac{\beta'_{2,y}}{2}\omega^2.
\end{equation}

The eigenvalues are found from a fourth-order characteristic polynomial equation, which can be written in a factorized form as
\begin{equation}
  \left[(\lambda+\alpha_x+i\Delta\beta'_1\omega)^2-f_x\right]\left[(\lambda+\alpha_y-i\Delta\beta'_1\omega)^2-f_y\right] = p,
\end{equation}
with $f_x = I_x^2-q_x^2$, $f_y = I_y^2-q_y^2$ and the XPM coupling parameter given by
\begin{equation}
  p = 4\sigma^2 I_xI_y(I_x-q_x)(I_y-q_y).
\end{equation}
Clearly, it is seen that the eigenvalues decouple and reduce to those of the scalar LLE in the limit of $\sigma \to 0$. We also note that the analysis can easily be extended to include higher order dispersion terms by suitably redefining the diagonal elements of the coefficient matrix.

The explicit solution of the characteristic equation is generally rather complicated. However, an important simplification occurs for the case of equal losses and zero GVM, i.e. $\alpha_x = \alpha_y = \alpha$ and $\Delta\beta'_1 \to 0$. This implies that the free spectral range of the two modes is equal, which can be accomplished by dispersion engineering, as it has been demonstrated in Ref.~\cite{Bao_2017}. The characteristic polynomial is then biquadratic and can easily be solved to give the eigenvalues
\begin{equation}
  \lambda = -\alpha \pm \sqrt{\frac{1}{2}(f_x+f_y) \pm \sqrt{p + \frac{1}{4}(f_x-f_y)^2}}.
  \label{eq:MI}
\end{equation}
We note that the eigenvalues are either real or complex conjugated. Furthermore, they are seen to approach their scalar counterparts for $|f_x-f_y| \gg p$, with a minimum separation for $f_x = f_y$. Neglecting the losses, we find that the fulfillment of either of the following three conditions will result in eigenvalues with a positive real part
\begin{equation}
  A_1:f_x+f_y > 0, \quad A_2:~p > f_xf_y, \quad A_3:~p < - \frac{1}{4}(f_x-f_y)^2.
  \label{eq:ineq1}
\end{equation}
Within these parametric regions we may thus expect an instability to develop, provided that the growth rate surpasses the absorption loss. The instabilities correspond to different eigenvectors (normal modes), whenever different combinations of eigenvalues are involved in the fulfillment of the conditions. The first instability $A_1$ occurs preferentially when both phase-matching conditions are satisfied simultaneously, i.e. $q_x = q_y = 0$. The second instability $A_2$ will generally involve the same eigenvalue as the first, but shows two asymptotic branches going off in individual directions specified by the conditions $q_x = 0$ and $q_y = 0$, respectively. Moreover, we find that the third instability $A_3$ depends on a complex conjugated pair of eigenvalues and has a branch along which $q_x + q_y = 0$ as the frequency increases. It can be shown that the $A_2$ and $A_3$ inequalities in Eq.~(\ref{eq:ineq1}) are mutually exclusive.

The phase-matching conditions can each be satisfied for either normal or anomalous dispersion. However, similar to the scalar case \cite{Hansson_2013}, for normal dispersion it is necessary that the detuning is sufficiently large to compensate for the nonlinear phase mismatch due to both SPM and XPM. This is in stark contrast to the case of an XPM coupled non-resonant system \cite{Agrawal_1987}. Instabilities of the $A_2$ and $A_3$ type can occur for the mixed case when the dispersion of one mode family is normal and the other is anomalous. In particular, we see that the phase-matching condition $q_x + q_y = 0$ may even become independent of frequency when the two modes have dispersion of equal magnitude but opposite sign. This suggests the possibility of an instability with a bandwidth that is limited only by higher orders of dispersion.

Another complementary reduction that shows the influence of GVM and unequal loss on the MI process can be accomplished by assuming symmetric modes with equal power, detuning and GVD, i.e. $I_x \to I_y \to I_0$ and $q_x \to q_y \to q_0$. In this case the eigenvalues simplify to
\begin{equation}
  \lambda = -\frac{1}{2}(\alpha_x+\alpha_y) \pm \sqrt{\left(f_0-\frac{1}{4}\zeta^2\right) \pm \sqrt{p_0 - f_0\zeta^2}},
  \label{eq:MI2}
\end{equation}
with $f_0 = I_0^2-q_0^2$, $p_0 = 4\sigma^2I_0^2(I_0-q_0)^2$ and $\zeta = 2\Delta\beta'_1\omega - i(\alpha_x-\alpha_y)$.
Here we find that an instability may occur for either of
\begin{equation}
  B_1:~f_0 > \frac{1}{4}\zeta^2, \quad B_2:~p_0 > \left(f_0+\frac{1}{4}\zeta^2\right)^2, \quad B_3:~p_0 < f_0\zeta^2.
\end{equation}
The asymptotic phase-matching conditions are in this case given by $q_0 = 0$ and $q_0 \pm \sqrt{(\zeta/2)^2+I_0^2} \approx q_0 \pm \zeta/2 = 0$. The presence of GVM makes it easy to satisfy the phase-matching conditions also for normal dispersion. As before, the $B_2$ and $B_3$ instabilities are mutually exclusive.

\begin{figure}[t]
\centerline{\includegraphics[width=1.1\linewidth]{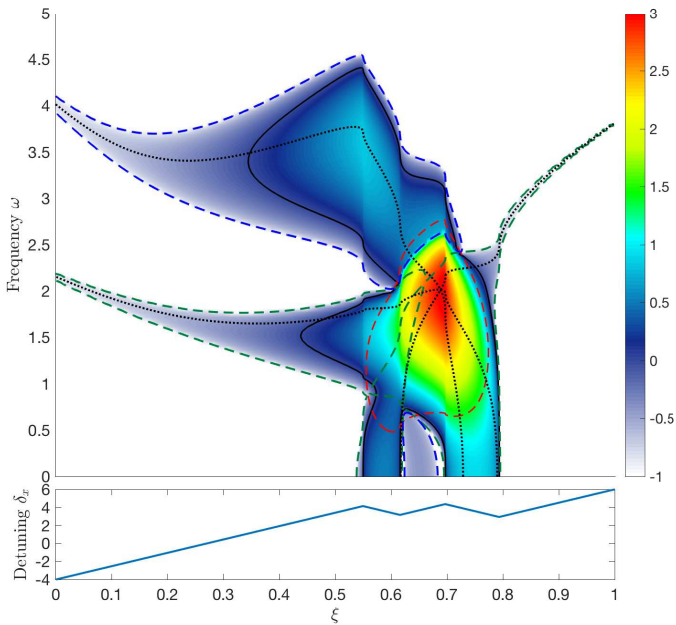}}
\caption{Top: Density plot of MI growth rate for parameters $P_x = P_y = 4$, $\alpha_x = \alpha_y = 1$ and $\delta_0 = 1.5$, with $\beta'_{2,x} = -2$ and $\beta'_{2,y} = 1$. The regions $A_1$, $A_2$ and $A_3$ that satisfy the MI conditions are denoted with red, green and blue dashed lines, respectively. The solid black line marks the contour for zero gain, while the dotted lines mark the phase-matching conditions. Bottom: Parameterization of detuning values for the tristable resonance curve.}
\label{fig:MI15}
\end{figure}

\begin{figure}[htb]
\centerline{\includegraphics[width=1.1\linewidth]{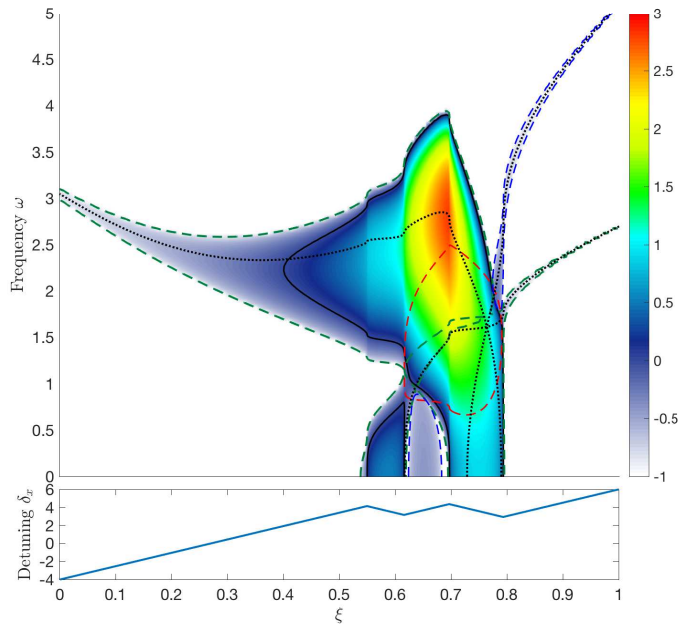}}
\caption{Density plot of MI growth rate for the same parameters as Fig.~\ref{fig:MI15}, except with $\beta'_{2,x} = -1$ and $\beta'_{2,y} = 2$. Here the $A_3$ type of instability is not observed, and it is the $A_2$ type that is the first to occur as the detuning increases.}
\label{fig:MI15b}
\end{figure}

To investigate the influence of MI in more detail, we consider some specific parameter conditions. To this end, we first parameterize the CW resonance curve by a continuous variable $\xi = \xi(\delta_x,I_x+I_y)$ that connects neighbouring values. $\xi$ corresponds to the detuning variable (x-axis) in a rotated coordinate frame where the CW solution in Fig.~\ref{fig:res} is single valued. This allows us to trace out the full curve by a single modified detuning parameter even in the case that the resonance is multistable. We next consider the MI growth rate along the resonance curve as a function of both $\xi$ and frequency $\omega$. Note that in the regions where $\xi$ maps to a detuning with negative slope the solutions are always unstable to CW perturbations. In Figs.~(\ref{fig:MI15}-\ref{fig:MI15b}) we show two examples of the MI growth rate $\textrm{Re}[\lambda]$ as predicted by Eq.~(\ref{eq:MI}). It can be seen that different types of instabilities, with separate ranges of unstable frequencies, are obtained depending on where the CW solution lies on the tilted resonance curve. The phase-matching conditions are also seen to be in good agreement with the asymptotic behavior for the variation of the MI gain as the frequency increases. They may consequently be utilized to make simple predictions about the instability conditions.

\begin{figure}[htb]
\centerline{\includegraphics[width=1.1\linewidth]{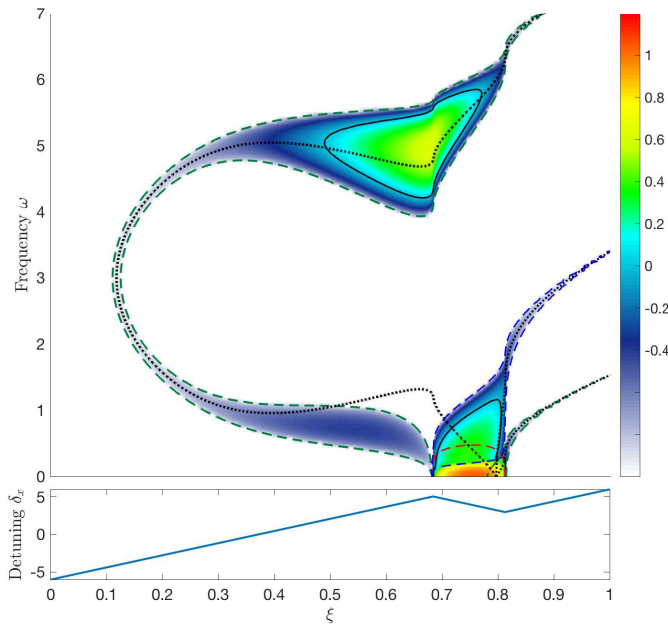}}
\caption{Top: Density plot of MI growth rate for $P_x = P_y = 3$, $\alpha_x = \alpha_y = 1$ and $\delta_0 = 0$, with normal dispersion $\beta'_{2,x} = \beta'_{2,y} = 1$ and GVM coefficient $\Delta\beta_1' = 3$. The regions $B_1$, $B_2$ and $B_3$ that satisfy the MI conditions are denoted with red, green and blue dashed lines, respectively. Bottom: Parameterization of bistable detuning values.}
\label{fig:MI0_GVM}
\end{figure}

Meanwhile, in Fig.~\ref{fig:MI0_GVM} we show another example for the complementary case of instabilities predicted by Eq.~(\ref{eq:MI2}). In particular, we consider a case with symmetric parameters for two mode families having normal dispersion. The CW solution is stable in the absence of GVM, but in its presence it is found to become modulationally unstable. Polarization mode coupling in combination with GVM may therefore allow for an alternative method for the generation of frequency combs in resonators characterized by normal dispersion. The phase-matching conditions are also here seen to be in good agreement with the asymptotic instability behavior. Note however, that Fig.~\ref{fig:MI0_GVM} does not represent a faithful frequency scan of the resonance, since the detuning difference $\delta_0$ will not be constant when the GVM is non-zero.

\section{Simulations of Comb Dynamics}

We now show numerical simulation results of frequency comb dynamics for the different types of instabilities predicted by the MI analysis. The simulations are performed by simultaneously solving Eqs.(\ref{eq:S1}-\ref{eq:S2}) using a split-step Fourier method, with the solution of the nonlinear step calculated using a 4th-order variable step-length Runge-Kutta algorithm. The temporal window for the simulations is assumed to have a fixed duration from $\tau \in [-5,5]$, which corresponds to a normalized free-spectral-range of $0.1$.

\begin{figure}[htb]
\begin{minipage}{\linewidth}
\centerline{\includegraphics[width=1.2\linewidth]{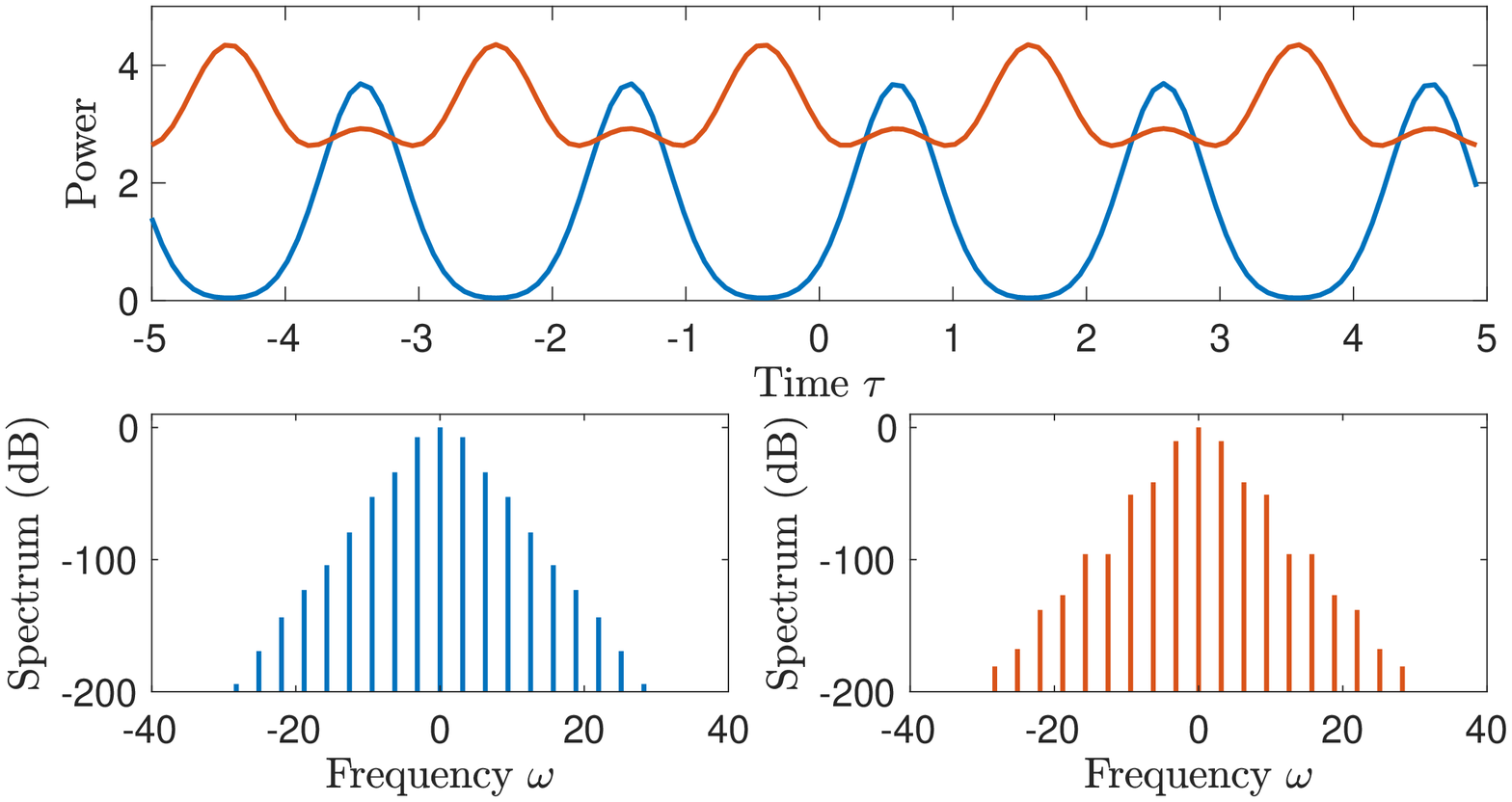}}
\end{minipage}
\begin{minipage}{\linewidth}
\centerline{\includegraphics[width=1.2\linewidth]{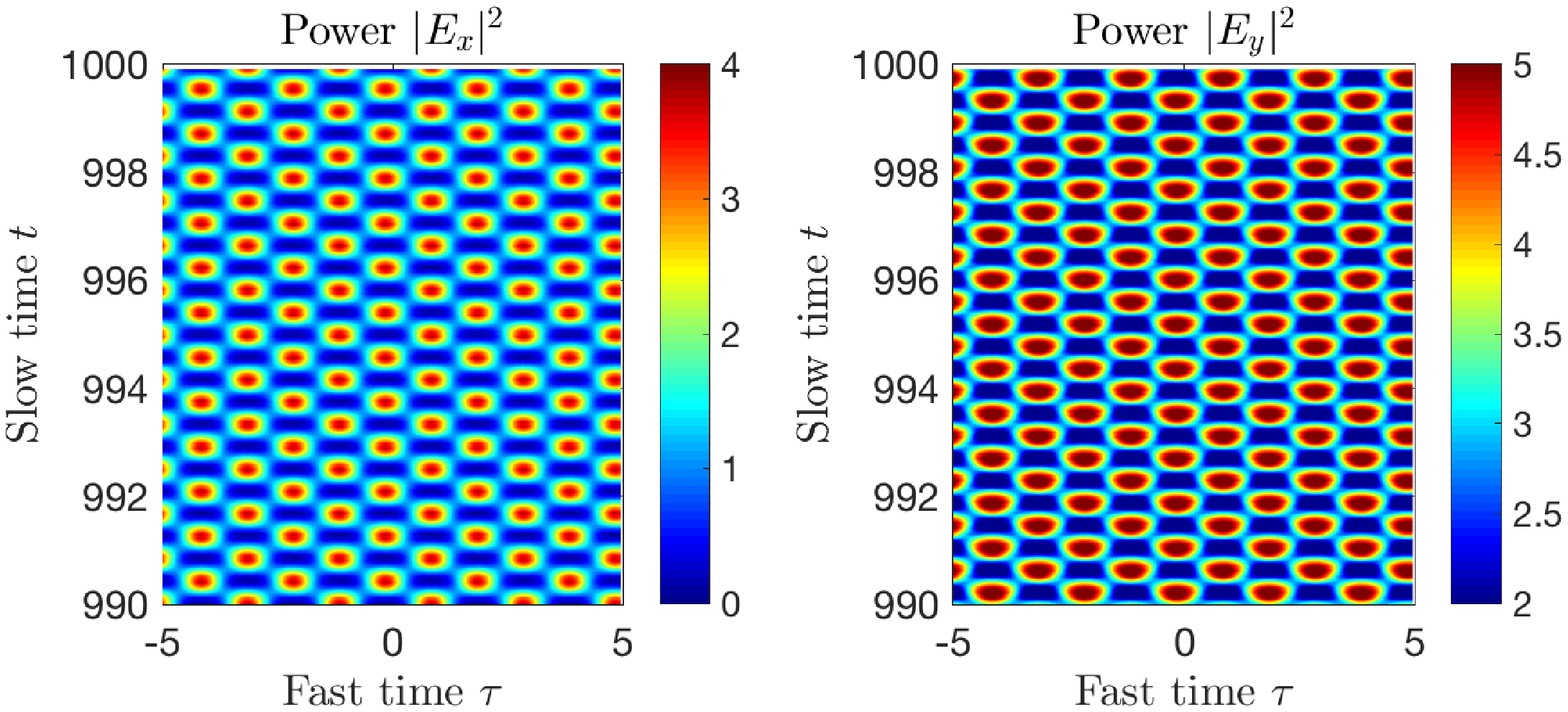}}
\end{minipage}
\caption{Top: Temporal intracavity power at $t = 1000$, with $|E_x|^2$ in blue and $|E_y|^2$ in orange. Comb generation occurs due to an instability of the $A_3$ type. Parameters as in Fig.~\ref{fig:MI15} with $\xi = 0.44$ ($\delta_x = 2.5$). Middle: Corresponding normalized spectrum on dB scale. Bottom: Temporal evolution showing periodicity of the vector state.}
\label{fig:run1}
\end{figure}

In the first case, we consider simulation parameters corresponding to Fig.~\ref{fig:MI15}, with a fixed detuning of $\xi = 0.44$ ($\delta_x = 2.5$). Here we expect an instability of the $A_3$ type with a pair of complex conjugated eigenvalues having a non-zero imaginary part. Such an instability can be convective \cite{Mussot_2008}, and involves a motion of the temporal pattern with respect to the group velocity at the resonance frequency of the pump. In fact, a translation is seen as the instability develops, however the final state is a group-velocity locked, temporally periodic pattern where the peak pulse intensity alternates between the two polarization components as shown in the bottom panel of Fig.~\ref{fig:run1}. This sort of dynamic pattern state has a spectrum where the comb lines oscillate in amplitude, and is not found in the scalar case where the eigenvalues are purely real.  

In the second case, we consider an identical CW resonance curve, but with group-velocity dispersion parameters having magnitudes as in Fig.~\ref{fig:MI15b} and a detuning of $\xi = 0.5$ ($\delta_x = 3.3$). The instability is now of the $A_2$ type and involves one real eigenvalue and a symmetric pair of growing modes. Here we observe the formation of a stationary pattern state consisting of a train of bright pulses for the x-component with anomalous dispersion, that are coupled with dark pulses (i.e. intensity dips) for the normally dispersive y-component. The y-component would here be stable against MI in the absence of coupling, but the development of modulations in the x-component results in a temporal variation of the intensity dependent nonlinear refractive index, which by necessity induces a frequency comb also in the orthogonally polarized y-component through the XPM coupling, c.f.~\cite{Bao_2017}.
\begin{figure}[htb]
\centerline{\includegraphics[width=1.2\linewidth]{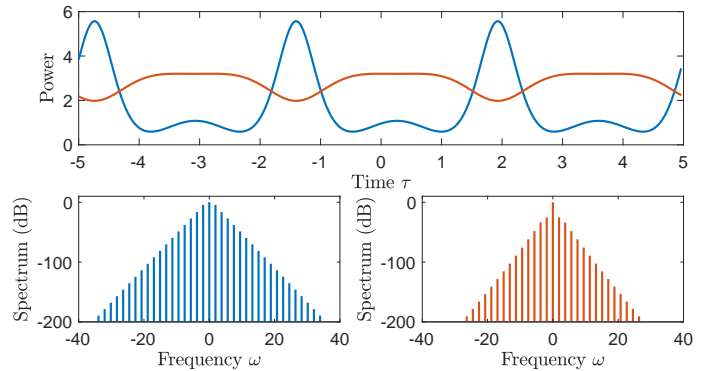}}
\caption{Top: Temporal intracavity power showing stationary pattern state at $t = 1000$ due to an instability of the $A_2$ type. The parameters are the same as in Fig.~\ref{fig:MI15b} with initial conditions given by the CW solution for $\xi = 0.5$ ($\delta_x = 3.3$). Bottom: Normalized spectrum on dB scale.}
\label{fig:run3}
\end{figure}

\begin{figure}[htb]
\centerline{\includegraphics[width=1.2\linewidth]{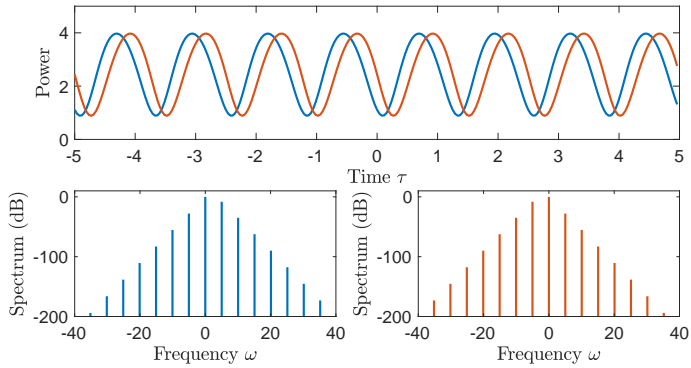}}
\caption{Top: Temporal intracavity power of stationary pattern state at $t = 1000$ resulting from normal dispersion comb generation with GVM. Parameters as in Fig.~\ref{fig:MI0_GVM} with $\xi = 0.6$ ($\delta_x = 4.5$). Bottom: Normalized spectrum on dB scale.}
\label{fig:run2}
\end{figure}

In the third case, we consider degenerate pump, detuning and GVM parameters for two modes with normal dispersion as shown in Fig.~\ref{fig:MI0_GVM}. To investigate the influence of GVM on the comb generation process, we assume that a temporal walk-off is present with a magnitude of $\Delta\beta_1' = 3$. The GVM allows an instability of the $B_2$ type to become phase-matched so that comb generation can occur. We find that the MI development results in the generation of coupled frequency combs and a stationary sinusoidal temporal pattern state with an asymmetric spectrum as shown in Fig.~\ref{fig:run2}.

\section{Conclusions}

The modulational instability of the intracavity field in microresonator devices is key to understanding the various nonlinear wave-mixing processes that lead to frequency comb generation. New phenomena are found to occur when the scalar description is generalized to take into account the vector nature of the electromagnetic field through the nonlinear coupling of orthogonal polarization modes. Linear mode interactions may also become important under various circumstances, although their characterization requires a more complicated analysis that is left for future work, c.f.~\cite{Guo_2017}.

Here we have shown that the evolution of the field components in the presence of nonlinear polarization mode coupling can be modelled by two coupled mean-field equations that generalize the Lugiato-Lefever equation. These equations admit a family of homogeneous steady-state solutions that can be characterized by the total power and relative detuning of the two modes. We have performed a stability analysis of these solutions, and demonstrated that the eigenvalues can be calculated explicitly for two complementary cases. The MI analysis reveals, in particular, that different eigenvalues can be responsible for the instability development, depending on the fulfillment of three simple phase-matching conditions. As the eigenvalues correspond to different normal modes, they can allow for the generation of frequency comb states that have no counterpart for the scalar case. Among them are group-velocity locked vector states which are temporally periodic on both the fast and the slow timescale.

An important characteristic of the nonlinear coupling of polarization modes is that it can allow for various alternative means of comb generation in different spectral regimes. The polarization degree of freedom may, e.g., permit MI to develop for normal dispersion, as is known for the case of polarization coupled fields in non-resonant structures such as optical fibers. Here, we have demonstrated such a possibility for the case of symmetric modes with a group-velocity mismatch. However, our analysis indicates that the phase-matching conditions will generally not be satisfied for modes with normal dispersion in the absence of GVM, except within a limited range of intracavity powers, in cases where the respective cavity detunings are sufficiently large.

\section*{Acknowledgements}

The research leading to these results has received funding from the European Union's Horizon 2020 research and innovation programme under the Marie Sklodowska-Curie grant agreement No GA-2015-713694. Additionally, T.H. and M.B. acknowledges funding by the Italian Ministry of University and Research (MIUR) (Project PRIN 2015KEZNYM - NEMO); S.W. acknowledges funding from the European Research Council (ERC) under the European Union's Horizon 2020 research and innovation programme (grant agreement No. 740355).


\begin{thebibliography}{1}
\bibitem{Kippenberg_2011} T.~J.~Kippenberg, R.~Holzwarth, and S.~A.~Diddams, ``Microresonator-based optical frequency combs,'' Science \textbf{332}, 555–559 (2011).
\bibitem{Pasquazi_2017} A.~Pasquazi, M.~Peccianti, L.~Razzari, D.~J.~Moss, S.~Coen, M.~Erkintalo, Y.~K.~Chembo, T.~Hansson, S.~Wabnitz, P.~Del’Haye, X.~Xue, A.~M.~Weiner, and R.~Morandotti, ``Micro-combs: A novel generation of optical sources,'' Physics Reports (2017).
\bibitem{Matsko_2005} A.~Matsko, A.~Savchenkov, D.~Strekalov, V.~Ilchenko, and L.~Maleki, ``Optical hyperparametric oscillations in a whispering-gallery-mode resonator: Threshold and phase diffusion,'' Phys. Rev. A \textbf{71}, 033804 (2005).
\bibitem{Hansson_2013} T.~Hansson, D.~Modotto, and S.~Wabnitz, ``Dynamics of the modulational instability in microresonator frequency combs,'' Phys. Rev. A \textbf{88}, 023819 (2013).
\bibitem{Coillet_2013} A.~Coillet, I.~Balakireva, R.~Henriet, K.~Saleh, L.~Larger, J.~M.~Dudley, C.~R.~Menyuk, and Y.~K.~Chembo, ``Azimuthal Turing patterns, bright and dark cavity solitons in Kerr combs generated with whispering-gallery-mode resonators,'' IEEE Photon. J. \textbf{5}, 6100409–6100409 (2013).
\bibitem{Qi_2017} Z.~Qi, G.~D’Aguanno, and C.~R.~Menyuk, ``Nonlinear frequency combs generated by cnoidal waves in microring resonators,'' J. Opt. Soc. Am. B \textbf{34}, 785 (2017).
\bibitem{LL_1987} L.~A.~Lugiato and R.~Lefever, ``Spatial dissipative structures in passive optical systems,'' Phys. Rev. Lett. \textbf{58}, 2209–2211 (1987).
\bibitem{Coen_2013} S.~Coen, H.~G.~Randle, T.~Sylvestre, and M.~Erkintalo, ``Modeling of octave-spanning Kerr frequency combs using a generalized mean-field Lugiato–Lefever model,'' Opt. Lett. \textbf{38}, 37–39 (2013).
\bibitem{Hansson_2015}  T.~Hansson and S.~Wabnitz, ``Frequency comb generation beyond the Lugiato–Lefever equation: multi-stability and super cavity solitons,'' J. Opt. Soc. Am. B \textbf{32}, 1259--1266 (2015).
\bibitem{Liu_2014} Y.~Liu, Y.~Xuan, X.~Xue, P.-H.~Wang, S.~Chen, A.~J.~Metcalf, J.~Wang, D.~E.~Leaird, M.~Qi, and A.~M.~Weiner, ``Investigation of mode coupling in normal-dispersion silicon nitride microresonators for Kerr frequency comb generation,'' Optica 1, \textbf{137} (2014).
\bibitem{Agrawal_1987} G.~P.~Agrawal, ``Modulation instability induced by cross-phase modulation,'' Phys. Rev. Lett. \textbf{59}, 880 (1987).
\bibitem{Wabnitz_1988} S.~Wabnitz, ``Modulational polarization instability of light in a nonlinear birefringent dispersive medium,'' Phys. Rev. A \textbf{38}, 2018–2021 (1988).
\bibitem{Rothenberg_1990} J.~E.~Rothenberg, ``Modulational instability for normal dispersion,'' Phys. Rev. A \textbf{42}, 682 (1990).
\bibitem{Savchenkov_2012} A.~A.Savchenkov, A.~B.~Matsko, W.~Liang, V.~S.~Ilchenko, D.~Seidel, and L.~Maleki, ``Kerr frequency comb generation in overmoded resonators,'' Opt. Express \textbf{20}, 27290–27298 (2012).
\bibitem{Cole_2017} D.~C.~Cole, E.~S.~Lamb, P.~Del’Haye, S.~A.~Diddams, and S.~B.~Papp, ``Soliton crystals in Kerr resonators,'' Nature Photonics \textbf{11}, 671–676 (2017).
\bibitem{Haelterman_1994} M.~Haelterman, S.~Trillo, and S.~Wabnitz, ``Polarization multistability and instability in a nonlinear dispersive ring cavity,'' J. Opt. Soc. Am. B \textbf{11}, 446–456 (1994).
\bibitem{Menyuk_2016} G.~D’Aguanno and C.~R.~Menyuk, ``Nonlinear mode coupling in whispering-gallery-mode resonators,'' Phys. Rev. A \textbf{93}, 043820 (2016).
\bibitem{Menyuk_2017} G.~D’Aguanno and C.~R.~Menyuk, ``Coupled Lugiato-Lefever equation for nonlinear frequency comb generation at an avoided crossing of a microresonator,'' Eur. Phys. J. D \textbf{71}, (2017).
\bibitem{Averlant_2017} E.~Averlant, M.~Tlidi, K.~Panajotov, and L.~Weicker, ``Coexistence of cavity solitons with different polarization states and different power peaks in all-fiber resonators,'' Opt. Lett. \textbf{42}, 2750 (2017).
\bibitem{Bao_2017} C.~Bao, P.~Liao, A.~Kordts, L.~Zhang, A.~Matsko, M.~Karpov, M.~H.~P.~Pfeiffer, G.~Xie, Y.~Cao, Y.~Yan, A.~Almaiman, Z.~Zhao, A.~Mohajerin-Ariaei, A.~Fallahpour, F.~Alishahi, M.~Tur, L.~Maleki, T.~J.~Kippenberg, and A.~E.~Willner, ``Orthogonally polarized Kerr frequency combs,'' arXiv:1705.05045 (2017).
\bibitem{Wang_2017} Y.~Wang, F.~Leo, J.~Fatome, M.~Erkintalo, S.~G.~Murdoch, and S.~Coen, ``Universal mechanism for the binding of temporal cavity solitons,'' Optica \textbf{4}, 855 (2017).
\bibitem{Agrawal} G.~P.~Agrawal, \textit{Nonlinear fiber optics, 5th Ed.} (Academic Press, New York, 2012).
\bibitem{Haelterman_1992} M.~Haelterman, S.~Trillo, and S.~Wabnitz, ``Dissipative modulation instability in a nonlinear dispersive ring cavity,'' Opt. Comm. \textbf{91}, 401–407 (1992).
\bibitem{Matsko_2012} A.~B.~Matsko, A.~A.~Savchenkov, V.~S.~Ilchenko, D.~Seidel, and L.~Maleki, ``Hard and soft excitation regimes of Kerr frequency combs,'' Phys. Rev. A \textbf{85}, (2012).
\bibitem{Mussot_2008} A.~Mussot, E.~Louvergneaux, N.~Akhmediev, F.~Reynaud, L.~Delage, and M.~Taki, ``Optical fiber systems are convectively unstable,'' Phys. Rev. Lett. \textbf{101}, (2008).
\bibitem{Guo_2017} H.~Guo, E.~Lucas, M.~H.~P.~Pfeiffer, M.~Karpov, M.~Anderson, J.~Liu, M.~Geiselmann, J.~D.~Jost, and T.~J.~Kippenberg, ``Intermode breather solitons in optical microresonators,'' Phys. Rev. X \textbf{7}, 041055 (2017).
\end{thebibliography}
\end{document}